\documentclass[11pt,epsf]{article}

\usepackage{epsfig}
\usepackage{amssymb}
\usepackage{graphicx}
\usepackage{color}
\usepackage{subfigure}

\begin{document}

\baselineskip 6mm
\renewcommand{\thefootnote}{\fnsymbol{footnote}}


\newcommand{\nc}{\newcommand}
\newcommand{\rnc}{\renewcommand}



\newcommand{\tcb}{\textcolor{blue}}
\newcommand{\tcr}{\textcolor{red}}
\newcommand{\tcg}{\textcolor{green}}


\def\be{\begin{equation}}
\def\ee{\end{equation}}
\def\ba{\begin{array}}
\def\ea{\end{array}}
\def\bea{\begin{eqnarray}}
\def\eea{\end{eqnarray}}
\def\nn{\nonumber\\}


\def\ct{\cite}
\def\la{\label}
\def\eq#1{(\ref{#1})}


\def\a{\alpha}
\def\b{\beta}
\def\g{\gamma}
\def\G{\Gamma}
\def\d{\delta}
\def\D{\Delta}
\def\ep{\epsilon}
\def\e{\eta}
\def\ph{\phi}
\def\Ph{\Phi}
\def\ps{\psi}
\def\Ps{\Psi}
\def\k{\kappa}
\def\l{\lambda}
\def\L{\Lambda}
\def\m{\mu}
\def\n{\nu}
\def\th{\theta}
\def\Th{\Theta}
\def\r{\rho}
\def\s{\sigma}
\def\S{\Sigma}
\def\ta{\tau}
\def\o{\omega}
\def\O{\Omega}
\def\pr{\prime}


\def\half{\frac{1}{2}}

\def\goto{\rightarrow}

\def\na{\nabla}
\def\grad{\nabla}
\def\curl{\nabla\times}
\def\div{\nabla\cdot}
\def\pa{\partial}

\def\bra{\left\langle}
\def\ket{\right\rangle}
\def\lb{\left[}
\def\lc{\left\{}
\def\ls{\left(}
\def\ln{\left.}
\def\rn{\right.}
\def\rb{\right]}
\def\rc{\right\}}
\def\rs{\right)}

\def\vac#1{\mid #1 \rangle}


\def\td#1{\tilde{#1}}
\def\check{ \maltese {\bf Check!}}


\def\Tr{{\rm Tr}\,}
\def\det{{\rm det}}


\def\bc#1{\nnindent {\bf $\bullet$ #1} \\ }
\def\ch {$<Check!>$ }
\def\ss {\vspace{1.5cm}}

\begin{titlepage}

\hfill\parbox{5cm} { }

\vspace{25mm}

\begin{center}
{\Large \bf Holographic Symmetry Energy of the Nuclear Matter}

\vskip 1. cm
  {Chanyong Park$^a$\footnote{e-mail : cyong21@sogang.ac.kr}}

\vskip 0.5cm

{\it $^a\,$Center for Quantum Spacetime (CQUeST), Sogang University, Seoul 121-742, Korea}\\

\end{center}

\thispagestyle{empty}

\vskip2cm


\centerline{\bf ABSTRACT} \vskip 4mm

\vspace{1cm}

We calculate the symmetry energy of the nuclear matter by using the bottom-up approach, so called hard wall model. To consider the nuclear matter, we introduce the isospin for u- and d-quarks. We find that in the hard wall model, the symmetry energy of the nuclear matter is proportional to the square of nucleon density. We also study the symmetry energy of the quark matter in the deconfining phase. Finally, we investigate the effect of the symmetry energy on the Hawking-Page transition and show that at the given quark density, the Hawking-Page transition temperature decreases due to the symmetry energy.

\vspace{2cm}


\end{titlepage}

\renewcommand{\thefootnote}{\arabic{footnote}}
\setcounter{footnote}{0}

\tableofcontents

\section{Introduction}

Knowing the density dependence of nuclear symmetry energy is an important issue understanding not only the structure of radioactive nuclei \cite{Brown:2000pd,Horowitz:2000xj,Furnstahl:2001un} but also nuclear astrophysics such as the neutron star and supernova
\cite{Bethe:1990mw,Lattimer:2000nx,Lattimer:2004pg}. It is well known that the nuclear symmetry energy at normal nuclear matter density is known to be around $30$MeV \cite{Myers:1966zz,Pomorski:2003mv} but the nuclear symmetry energy in other densities is poorly known \cite{Li:1997px,Chen:2004si}.
Although the lattice QCD is a powerful method to understand the strongly interacting QCD, there are still many problems regarding the dense medium effect. 
The AdS/CFT correspondence \cite{Maldacena:1997re,Witten:1998qj,Witten:1998zw} can shed light on understanding various aspects of the dense medium effect. Recently, by using this AdS/CFT correspondence the nuclear symmetry energy was investigated in the top-down approach \cite{Kim:2010dp}. In this paper, we will investigate the nuclear symmetry energy in the bottom-up approach, so called hard wall model, and the effect of it on the deconfiment phase transition of the QCD-like gauge theory.

In the original hard wall model \cite{Erlich:2005qh}, where an IR cut off was introduced to explain the confining behavior, various meson spectra and decay constants were investigated. 
After this original work, there were various extensions to improve the model, like the soft wall model \cite{Karch:2006pv} or top-down approaches in the string theory \cite{arXiv:hep-th/0412141,arXiv:hep-th/0611021}, and to include the gravitational back reaction of various bulk fields corresponding to the gluon condensate \cite{hep-th/9902155} or the chiral condensate \cite{Erlich:2005qh,hep-th/0311270}. By considering the Hawking-Page transition between the thermal AdS (tAdS) space and the Schwarzschild AdS black hole, the deconfiment phase transition of the dual gauge theory was explained in the zero quark density limit \cite{Herzog:2006ra}. This calculation was generalized to the case having nonzero quark density \cite{Domokos:2007kt,Kim:2007em,Lee:2009bya}. To describe the nonzero quark density, the time component of the vector gauge field, whose dual operator in the dual gauge theory corresponds to the quark density, was introduced, which modifies the background geometry. In the deconfining phase, the geometry of the dual gravity is described by a Reinssner-Nordstr\"{o}m AdS (RNAdS) black hole. The geometry corresponding to the confining phase should be a non-black hole solution. The solution satisfying the Einestein and Maxwell equations is given by the thermal charged AdS (tcAdS) space, which has a singularity at the center. To describe the confining behavior we should introduce an IR cut off, which can also resolve the singularity problem at the center of the tcAdS space. By comparing the free energies of the tcAdS and RNAdS, the deconfinement phase transition depending on the temperature and quark density was investigated and it was also shown that the resulting shape of the phase diagram is very similar to the expectation of the phenomenology \cite{Lee:2009bya}. In addition, the screening effect of the heavy quarkonium \cite{hep-th/0605182,hep-th/0605158,hep-ph/0607062,arXiv:0809.1336} and the spectra of various light mesons were also investigated in the tcAdS background \cite{Lee:2009bya}. The similar study was also done in the soft wall model \cite{hep-th/0603170}.
 
In this paper, we further generalize the above set-up by introducing the isospin. We first consider the U(2) flavor symmetry group, whose diagonal group elements give the information for the quark density and isospin of quarks of the dual gauge theory.  Using these informations, we can describe the quarks of the dual gauge theory as u- or d-quarks. In the confining phase, the number of u- and d-quarks can be easily identified with the number of nucleons, proton and neutron. If the number of u- and d-quarks are different, the free energy of this system has three parts. The first depends only on the IR cut off, which is independent of the quark density. The second is proportional to the total quark density. These two parts are irrelevant with the number difference between u- and d-quarks (or proton and neutron). The last one is proportional to the number difference between proton and neutron. The coefficient of this number difference in the free energy is usually called the symmetry energy. We found that the symmetry energy of the nuclear matter is proportional to the square of the nuclear density. 

We also investigate the similar symmetry energy in the deconfining phase. In the deconfining phase
corresponding to the RNAdS geometry, there exist free quarks instead of nucleons. We  calculate the free energy of this system, which contains a term proportional to the number difference between u- and d-quarks. We identify this energy with the symmetry energy of the quark matter. In general, the symmetry energy of the quark matter is a very complicated function of the temperature and quark density. In the high temperature and high density regime, we find that the symmetry energy of quark matter is proportional to the 
square of the total quark density and to the inverse square of the temperature. 

Finally, we investigate the effect of this symmetry energy on the Hawking-Page transition and show that the symmetry energy decreases the Hawking-Page transition temperature at the given quark density. 

The rest part is as follows: In Sec. 2, we explain our set-up and background solutions. In Sec. 3, the free energy of the nuclear matter is calculated and the nuclear symmetry energy is extracted from the nuclear free energy. In Sec. 4, we also calculate the symmetry energy of the quark matter. In Sec. 5, by comparing the results in Sec. 3 and Sec. 4, we investigate the Hawking-Page transition depending and the effect of the symmetry energy, Finally, we finish our work with concluding remarks in Sec. 6.

\section{Nuclear matter in the hard wall model}

In this section, we will consider the gravity theory dual to the nuclear matter following the AdS/QCD
correspondence. To describe the nuclear matter in the confining phase
or quark matter in the deconfing phase, we consider the following Euclidean
Einstein-Yang-Mills theory
\be \la{Act:org}
S = \int d^5 x \sqrt{G} \lb \frac{1}{2 \k^2} \ls - {\cal R} + 2 \L \rs + \frac{1}{4 g^2}
\Tr F_{MN} F^{MN} \rb ,
\ee
where $F_{MN} = \pa_M A_N - \pa_N A_M - i g \lb A_M , A_N \rb$. In the above,
$A_M = A_M^a T^a$ and $T^a$, where $a$ runs from $0$ to $N_f^2-1$, are the gauge field
and generator of the $U(N_f)$ flavor symmetry group.
To describe proton and neutron, we concentrate on the case having two flavors $N_f=2$.  For the $U(2)$ flavor symmetry,
the generators are proportional to the identity matrix ${\bf 1}$ and Pauli matrices $\s^{\td{a}}$
\be
T^0 = \frac{\bf 1}{2}  \ \ {\rm and} \ \ T^{\td{a}}= \frac{\s^{\td{a}}}{2}  \ \ (\td{a}=1,2,3) ,
\ee
where we use the normalization $\Tr T^a T^b = \half \d^{ab}$ ($a, b = 0,1,2,3$).
Usually,  the boundary values of the diagonal elements of the bulk gauge field, $A_0^0$ and $A_0^3$, have the definite meanings corresponding to the quark
and isospin chemical potential respectively. So  it is sufficient
to turn on $A_0^0$ and $A_0^3$ to describe proton and neutron, where the lower and upper
indices are for the space-time and flavor group respectively. In this set-up, the off-diagonal components
of the bulk gauge field can be considered as fluctuations describing the mesons of the dual
gauge theory.
Introducing
\be	\la{def:gaugept}
A^u_M \equiv  \ls A^0_M + A^3_M \rs \quad {\rm and} \quad
A^d_M \equiv  \ls A^0_M - A^3_M \rs,
\ee
the time component of the redefined gauge fields can describe u- and d-quark. 
In addition, the action  \eq{Act:org}  can be rewritten in terms of u- and d-quark as
gauge fields
\be	\la{act:nucl}
S = \int d^5 x \sqrt{G} \lb \frac{1}{2 \k^2} \ls - {\cal R} + 2 \L \rs + \frac{1}{16 g^2}
 F^u_{MN} F^{uMN}   + \frac{1}{16 g^2}  F^d_{MN} F^{dMN} \rb .
\ee
Here, since we turn on elements of the Cartan subgroup only, the gauge group
is reduced from $U(2)$ to $U(1) \times U(1)$ and the field strength of them becomes one for
the Abelian group
\be
F^{\a}_{MN} = \pa_M A^{\a}_N - \pa_N A^{\a}_M ,
\ee
where ${\a}$ implies u and d. From this action, the Einstein and Maxwell equation read off
\bea	    \la{eq:EM}
{\cal R}_{MN} - \half g_{MN} {\cal R} + g_{MN} \L &=&  \frac{\k^2}{4 g^2}  \sum_{i=u,d}
\lb g^{PQ}   F^{\a}_{MP} {F^{\a}_{NQ}} - \frac{1}{4} g_{MN} F^{\a}_{PQ} F^{{\a} PQ}\rb , \nn
0 &=& \frac{1}{\sqrt{G}} \pa_M \sqrt{G} g^{MP} g^{NQ} F^{\a}_{PQ} ,
\eea
which is the usual equation of motion for the Einstein-Maxwell theory having two U(1) charges.
So we can easily expect that the most general solution is given by the Reissner-Nordstr\"{o}m AdS (RNAdS) black hole including two charges, whose metric form is 
\be	\la{met:anz}
ds^2 = \frac{R^2}{z^2} \ls f(z) d t^2 + \frac{1}{f(z)} dz^2 + d \vec{x}^2 \rs .
\ee
To describe u- and d-quark, we turn on the time-component of gauge field depending only on
$z$-coordinate 
\be
A^{\a}_0 = A_{\a} (z)  \ \ {\rm and } \ \ A^{\a}_{i} = 0 \ \ ({\a}=u,d \ {\rm and} \ i=1,2,3)
\ee
where $i$ implies the spatial directions.
Then, the solution of the Maxwell equation is given by
\be	\la{sol:gauge}
A_{\a} (z) = i \ls \m_{\a} - Q_{\a}  \  z^2 \rs .
\ee
In the dual gauge theory side, $\m_{\a}$ and $Q_{\a}$ can be interpreted as the chemical
potential and the number density of u- or d-quark.
Inserting \eq{met:anz} and \eq{sol:gauge} into the Einstein equation, the black hole factor
$f(z) = f_{RN} (z)$
is given by
\be	 \la{met:RNAdS}
f_{RN} (z) = 1 - m z^4 + \ls {q_u}^2 + {q_d}^2 \rs z^6 ,
\ee
where $q_u$ and $q_d$ correspond to two black hole charges and $m$ is the black hole mass.
Notice that $q_{\a}$ and $Q_{\a}$ are not independent parameters, which satisfy
\be 	\la{rel:para}
q_{\a}^2  = \frac{\k^2}{6 g^2 R^2}  Q_{\a}^2  .
\ee
This RNAdS black hole is the most general solution and corresponds to the deconfining phase
of the dual gauge theory. To describe the nuclear matter, we should find another
geometric solution corresponding to the confining phase, which is not a black hole solution.
The non-black hole solution, which we call thermal charged AdS (tcAdS), is given by
\be
f(z) = f_{tc} (z) = 1 + \ls {q_u}^2 + {q_d}^2 \rs z^6 .
\ee
with the same relations in \eq{sol:gauge} and \eq{rel:para}, which also satisfies the
Einstein and Maxwell equations in \eq{eq:EM}.  For the tcAdS space to describe the confining
behavior, we should add an IR cut off at $z_{IR}$, which is called the hard wall model.
Then, the allowed range of $z$ is from $z_{IR}$ to $z=0$, where $z=0$ corresponds to the
boundary.
Although the tcAdS solution has a singularity at the center ($z \to \infty$),  we can safely use the tcAdS space in the hard wall model because the IR cut off prevents all bulk quantities from approaching to this singular point.
The basic degrees of freedom in the confining phase are not quarks but nucleons, so we
need to rewrite the u- and d- gauge fields in terms of ones for nucleons. Anyway, for later convenience we continue to use the u- and d-quark notations and rewrite them later in terms
of nuclear quantities.

\section{The holographic free energy of the nuclear matter}

In this section, we will investigate the thermodynamic energy of the dual gauge theory.
Depending on the boundary condition of the bulk gravity theory, the dual gauge theory can be described by a grand canonical or canonical ensemble.  If we impose the Dirichlet
boundary condition to the bulk gauge field, which fix the chemical potential of the dual gauge
theory, the corresponding thermodynamic energy is described by the grand potential of
the grand canonical ensemble. On the contrary, the Neumann boundary condition fixes the
quark number density, so it gives the free energy describing the canonical ensemble. These two
thermodynamic free energies are related  by the Legendre transformation. Here, we will study
these thermodynamic free energies by using the holographic calculation.

First, we study the grand potential. Imposing the Dirichlet boundary condition at the
boundary $z=0$
\be
A_{\a} (0) = i \m_{\a}   \ \ \ ({\a} = u, d) ,
\ee
the on-shell gravity action of the tcAdS, in which we insert an IR cut off at $z_{IR}$, becomes
\bea
S_{tc}^D &=& \int_0^{\b_{tc}} d t \ \int d^3 x \ \int_{\ep}^{z_{IR}} d z \ \sqrt{G} \lb \frac{1}{2 \k^2} (- {\cal R} + 2 \L)
+ \frac{1}{16 g^2}  \ls F^u_{MN} F^{uMN} + F^d_{MN} F^{dMN} \rs \rb \nn
&=& \frac{R^3 V_3  \b_{tc}}{\k^2} \ls \frac{1}{\ep^4} - \frac{1}{z_{IR}^4} - \frac{\k^2 z_{IR}^2}{6 g^2 R^2}
(Q_u^2 + Q_d^2)  \rs ,
\eea
where $V_3$, $\ep$ and $\b_{tc}$ means the three-dimensional volume, the UV cut off and
the periodicity in the Euclidean time coordinate. Taking the $\ep \to 0$ limit makes the above on-shell
action divergent, so we should renormalized it for the finiteness of the on-shell action.
Although there are several ways to renormalize it, here we adjust a background subtraction method in which
we subtract the background effect from the above. Since the asymptotic geometry of our model
is a five-dimensional Euclidean AdS or thermal AdS (tAdS), we remove the contributions of tAdS space from the above. The tAdS is
described by the following action
\be
S_{t} = \int_0^{\b_t} d t\int  d^3 x  \int_{\ep}^{\infty} d z\frac{1}{2 \k^2}   \ls - {\cal R} + 2 \L \rs,
\ee
whose solution satisfying the Einstein equation is given by
\be
ds^2 = \frac{R^2}{z^2} \ls d t^2 + dz^2 + d \vec{x}^2 \rs .
\ee
Then, the on-shell action for the tAdS becomes
\be
S_{t} = \frac{R^3 V_3 \b_t}{\k^2} \frac{1}{\ep^4} ,
\ee
where we use a different time periodicity $\b_t$, which can be determined in terms of $\b_{tc}$
by requiring the same circumferences of tAdS and tcAdS at the UV cut off
\be
\b_t = \b_{tc} + {\cal  O} (\ep^6) .
\ee
Notice that the correction term $ {\cal  O} (\ep^6)$ in the above does not give any effect as $\ep \to 0$.
So, the renormalized action with the Dirichlet boundary condition at the UV cut off is given by
\be		\la{onactD}
\bar{S}_{tc}^D =  \frac{R^3 V_3 \b_{tc}}{\k^2}  \ls  - \frac{1}{z_{IR}^4} - \frac{\k^2  z_{IR}^2}{6 g^2 R^2}
(Q_u^2 + Q_d^2) \rs.
\ee
The above on-shell action is related to the grand potential of the boundary theory
\be
\O_{tc} = \frac{\bar{S}_{tc}^D}{\b_{tc}} ,
\ee
which is function of $T$ and $\m_{\a}$. So $Q_{\a}$ in \eq{onactD} should be considered as a function of $\m_{\a}$.  To determine the relation between $Q_{\a}$ and $\m_{\a}$, we first assume
the following relation
\be \la{ass:rel}
Q_{\a} = c \m_{\a}  ,
\ee
where $c$ is an arbitrary constant. In general, we can choose $Q_\a$ as an arbitrary function of
$\m_\a$ but, as will be shown later, only the above form provides the well-defined Legendre
transformation. From the thermodynamic relation, the total number of quark is
given by
\be \la{numnuc}
N_{\a} = - \frac{\pa \O}{\pa \m_\a} =  \frac{ R  V_3 z_{IR}^2 }{3 g^2} \ c^2 \m_{\a} .
\ee
Therefore, after the Legendre transformation, the free energy becomes
\be	\la{res:free}
F_{tc} = \O + \sum_{{\a}=u,d} \m_{\a} N_{\a} = \O +  \frac{ R  V_3 z_{IR}^2 c }{3 g^2}    \sum_{{\a}=u,d}
\m_{\a}  Q_{\a} ,
\ee
in which the fundamental variable is not $\m_\a$ but $Q_\a$.

As shown in Ref. , this free energy can be calculated in a different way. If we impose the
Neumann boundary condition to the gravity action instead of the Dirichlet boundary
condition,  since the Neumann boundary condition fixes the quark density, the on-shell action
is proportional to the free energy of the boundary theory. To impose the Neumann boundary
condition, we should add a boundary term, which fixes the charge $Q_\a$, to the original action \eq{act:nucl}.  Vanishing of the action variation with respect to the gauge fields with
the Neumann boundary condition determines the additional boundary terms as
\bea	\la{act:boundary}
S_{B} &=& - \sum_{{\a} = u, d} \frac{1}{4 g^2} \int_{\pa M} d^4 x \sqrt{G^{(4)}} \ n^M g^{LN} A_L F_{MN} \nn
&=& \frac{ R V_3  }{2 g^2} \b_{tc} \sum_{\a=u,d} \m_\a Q_\a ,
\eea
where $G^{(4)}$ is the determinant of the boundary metric and the unit normal vector 
is given by $n^M = \lc 0,0,0,0, - \frac{z}{R} \sqrt{f_{tc} (z)} \rc$.
For the well-defined Legendre transformation, this additional boundary term after divided by $\b_{tc}$ should be same as the sum of $\m_{\a} N_\a$ in \eq{res:free}. This requirement fixes the constant $c$ to be
\be \la{res:rel}
c =  \frac{3}{2 z_{IR}^2}.
\ee
Substituting this relation into \eq{numnuc}, we finally obtain
\be \la{res:num}
N_{\a} = \frac{ R V_3}{2 g^2} Q_{\a} .
\ee
The value of $g^2$ can be determined by several ways. By comparing the current-current
correlation function of this model with the result of the phenomenological model, $g^2$
was fixed by $1/g^2 = N_c/(12 \pi^2 R)$ \cite{Erlich:2005qh,DaRold:2005ju}. Also, by comparing the D3-D7 brane set-up, the different value of $1/g^2 = N_c / (4 \pi^2 R)$
was used \cite{Kim:2007em,Lee:2009bya}. In all case, $1/g^2$ is proportional to the number of color. In this paper, for the
convenience, we choose $g^2 = R/2 N_c$, where $N_c$ is the number of color. Then, from
\eq{res:num} $Q_\a$ can be interpreted as the number density of one color quark.

The free energy density of the canonical ensemble becomes
\be
\frac{F_{tc} }{V_3} = -\frac{R^3}{\k^2}  \frac{1}{z_{IR}^4} +\frac{2 g^2 z_{IR}^2}{3 R V_3^2} \ls N_u^2 + N_d^2 \rs .
\ee
As mentioned previously, the fundamental elements of the confining phase are not quarks
but nucleons. Since the number of particles are preserved in the canonical ensemble, the free energy
can be without loss of generality rewritten in terms of the number of proton $N_p$ and neutron $N_n$ by using the fact that proton (or neutron) is composed of two u-quarks and one d-quark (or one u-quark and two d-quarks)
\be
N_u = 2 N_p + N_n  \ \ {\rm and } \ \ N_d = N_p + 2 N_n .
\ee
If we denote the total number of nucleons by $N = N_p + N_n$, the free energy density
can be written as
\be		\la{res:freeconf}
\frac{F_{tc} }{V_3} = - \frac{R^3}{\k^2} \frac{1}{z_{IR}^4} + \frac{3 g^2 z_{IR}^2}{ R} \r^2
+  \frac{g^2 z_{IR}^2}{3 R} \r^2  \ls \frac{N_p - N_n}{N} \rs^2
\ee
where $\r = N/V_3$ means the nucleon number density. Here, the coefficient of the last term,
which describe the contributions coming from the number difference between
proton and neutron, is called the symmetry energy $E_s$
\be
\frac{E_s}{V_3} = \frac{g^2 z_{IR}^2}{3 R} \r^2 .
\ee
Recently, it was shown that the symmetry energy is proportional to $\r^{1/2}$
in the top-down approach, especially D4-D6 brane set-up, numerically \cite{Kim:2010dp}. In the hard wall model, we find that
the symmetry energy is proportional to $\r^2$ with analytic calculation.

\section{ The holographic free energy of quark matter}

In this section, we will investigate the holographic free energy of quark matter composed of
two different quarks, $u$- and $d$-quarks. To investigate the quark matter holographically we should take into account the RNAdS black hole geometry. Though the RNAdS black hole geometry can be usually characterized by the black hole mass and charges, more fundamental quantities in the dual gauge theory are the temperature and quark density in the canonical ensemble (or the chemical potential in the grand canonical ensemble). So it is more convenient to rewrite the black hole mass in terms of the temperature and quark density. From the definition of the balck hole horizon $f_{RN} (z_h) = 0$, the black hole mass can be rewritten in terms of the temperature and the black hole horizon as
\be
m = \frac{1}{z_h^4} + (q_u^2 + q_d^2) z_h^2 ,
\ee
where the black hole charges $q_u$ and $q_d$ are directly related to the quark density. Inserting this relation to the definition of the Hawking temperature, we finally obtain
\be		\la{def:temp}
T_{RN} \equiv \frac{1}{\b_{RN}} = \frac{1}{z_h} \ls 1 - \half (q_u^2 + q_d^2) z_h^6 \rs ,
\ee
in which we can regards $z_h$ as a function of the temperature and quark density (or chemical potentials).

At first, we calculate the grand potential of the dual gauge theory for which we should impose the Dirichlet boundary condition at the boundary $A_{\a} = i \m_{\a}$. Then, the gravity one-shell action as a function of $T_{RN}$ and $\m_{\a}$ becomes
\be
S^D_{RN} = \frac{R^3 V_3}{\k^2} \b_{RN} \ls \frac{1}{\ep^4} - \frac{1}{z_h^4} - \frac{\k^2}{6 g^2 R^2} (Q_u^2 + Q_d^2) z_h^2 \rs,
\ee
where $\ep$ implies the UV cut off and $Q_{\a}$ is a function of the chemical potential $\m_{\a}$. To determine $Q_{\a}$, we should impose the regularity condition of $A_{\a}$ at the horizon, which is satisfied for $A_{\a} (z_h) = 0$. From this regularity condition, the charge $Q_{\a}$
can be reexpressed as a function of the chemical potential
\be	 	\la{rel:qm}
Q_{\a} = \frac{\m_{\a}}{z_h^2} .
\ee
In terms of the temperature and the chemical potentials, the on-shell gravity action is given by
\be
S^D_{RN} = \frac{R^3 V_3}{\k^2} \frac{1}{ T_{RN}} \ls \frac{1}{\ep^4} - \frac{1}{z_h^4} - \frac{\k^2}{6 g^2 R^2} (\m_u^2 + \m_d^2) \frac{1}{z_h^2} \rs ,
\ee
where $z_h$ is a function of the temperature and the chemical potentials. Here, when $\ep \to 0$ the above grand potential diverges, so that we should renormalized it.
By subtracting the on-shell gravity action of the background tAdS, we can find the renormalized
action related to the grand potential $\O$
\be	\la{grandpot}
\bar{S}_{RN} \equiv   \frac{\O_{RN} }{ T_{RN}} = \frac{V_3 R^3}{\k^2} \frac{1}{T_{RN}} \ls - \frac{1}{2 z_h^4} - \frac{\k^2}{12 g^2 R^2}   (\m_u^2 + \m_d^2) \frac{1}{z_h^2} \rs ,
\ee
where we use the fact that the temperature $T_t$ of the tAdS space is related to $T_{RN}$ of the RNAdS black hole
\be
\frac{1}{T_t} = \ls 1 - \half m \ep^4 \rs \frac{1}{T_{RN}} + {\cal O} (\ep^6) .
\ee
From the grand potential in \eq{grandpot}, the particle number can be calculated by using the thermodynamic relation $N_{\a} = - \frac{\pa \O}{\pa \m_{\a}}$. After the tedious calculation,
the quark number is given by
\be
N_{\a} = \frac{V_3 R}{2 g^2} Q_{\a} ,
\ee
which is the same form as one in \eq{res:num} calculated in the confining phase. Via a Legendre transformation the free energy in the canonical ensemble is given by
\be		\la{Res:FreeDeconf}
F_{RN}  \equiv \O + \sum_{\a=u,d} \m_{\a} N_{\a} = \frac{V_3 R^3}{\k^2} \ls - \frac{1}{2 z_h^4} + \frac{5 \k^2}{12 g^2 R^2} (Q_u^2 + Q_d^2) z_h^2 \rs ,
\ee
where $z_h$ is given by a function of $T_{RN}$ and the quark number density $Q_{\a}$. Notice that we can also reproduce the same free energy by calculating the on-shell gravity action together with the Neumann boundary condition, as shown in the previous section, instead of the Dirichlet boundary condition. In terms of the total quark density $Q = Q_u + Q_d$
and density difference $D = Q_u - Q_d$, the black hole horizon $z_h$ as a function of $T_{RN}$, $Q$ and $D$ can be determined from
\be
T_{RN} = \frac{1}{\pi z_h} \ls 1 - \frac{\k^2}{24 g^2 R^2} (Q^2 + D^2) z_h^6 \rs ,
\ee
and the free energy is rewritten as
\be
\frac{F}{V_3} = \frac{ R^3}{\k^2} \ls  - \frac{1}{2 z_h^4} + \frac{5 \k^2}{24 g^2 R^2} (Q^2 + D^2) z_h^2 \rs .
\ee

In the high quark density $Q \gg 1$ and the high temperature $z_h^6 \ll \frac{24 g^2 R^2}{\k^2 (Q^2 + D^2)}$ limit, the temperature can be approximated with $T_{RN} \sim 1/(\pi z_h)$, so the
free energy becomes
\be
\frac{F}{V_3} \approx \frac{R^3}{\k^2} \ls  - \frac{\pi^4 T_{RN}^4}{2} + \frac{5 \k^2  }{24 \pi^2  g^2 R^2}
\frac{1}{T_{RN}^2} Q^2+ \frac{5 \k^2  }{24 \pi^2 g^2 R^2}
\frac{Q^2}{ T_{RN}^2} \frac{D^2}{Q^2}  \rs .
\ee
From this result, we can also define the symmetry energy of the quark matter which is given by the coefficient of $\frac{D^2}{Q^2}$ via taking the analogy to the symmetry energy in the nuclear medium. Then, we can see that in the high temperature and high quark density limit, the symmetry energy of the quark matter becomes
\be
\frac{E_s}{V_3} \approx \frac{5 R  }{24 \pi^2 g^2 } \frac{Q^2}{ T_{RN}^2} .
\ee
From this, we can see that the symmetry energy of the quark matter is proportional to the square of the quark density and the inverse square of the temperature.

\section{Hawking-Page transition in the nuclear medium}

Now, we consider the Hawking-Page transition, which is identified with the deconfinement 
phase transition of the dual QCD-like theory. If we ignore the isospin of the bulk gauge field by turning off $A^3_0$ in \eq{def:gaugept}, there is no difference between $u$- and $d$-quarks, so that  $u$- and $d$-quarks have the same chemical potential energy and the number density.  In this case, the free energies of the confining and deconfining phase
simply reduce to ones in Ref. \cite{Lee:2009bya} by setting  $N_p = N_n$ in \eq{res:freeconf}
and $Q_u = Q_d$ in \eq{Res:FreeDeconf} and their Hawking-Page transition was 
fully investigated in Ref. \cite{Lee:2009bya} where the phase diagram expected by the phenomenology was obtained.
In this paper, we will investigate the effect of the symmetry energy caused by the isospine on the Hawking-Page transition.

In terms of the quark number density, the renormalized action of tcAdS and RNAdS black hole with the Neumann boundary condition are given by
\bea		\la{res:actions}
\bar{S}_{RN}^N
&=& \frac{R^3 V_3}{\k^2} \b_{RN} \ls - \frac{1}{2 z_h^4} + 
\frac{5 \k^2}{24 R^2 g^2} (Q^2 + D^2) z_h^2 \rs , \nn
\bar{S}_{tc}^N
&=& \frac{R^3 V_3}{\k^2} \b_{tc} \ls - \frac{1}{z_{IR}^4} + 
\frac{\k^2}{12 R^2 g^2} (Q^2 + D^2) z_{IR}^2 \rs ,
\eea
where $Q$ and $D$ are total quark number density and the density difference between $u$- and $d$-quarks respectively. By matching the Euclidean time circumferences of two geometries, we can
reexpress $\b_{tc}$ in terms of $\b_{RN}$ at the UV cut off, $\ep$,
\be		\la{rel:temps}
\b_{tc} = \b_{RN} + {\cal O} (\ep^4)
\ee
Because the actions in \eq{res:actions} have no divergence, the second term in \eq{rel:temps}
does not play any role for $\ep \to 0$. As a result, the action difference between these two
geometries is given by
\bea		\la{eq:hp}
\D S &\equiv& \bar{S}_{RN}^N - \bar{S}_{tc}^N \nn
&=& \frac{R^3 V_3}{\k^2} \b_{RN} \lb  \frac{1}{z_{IR}^4} - \frac{1}{2 z_h^4}
+ \frac{\k^2}{24 R^2 g^2} ( 5 z_h^2 - 2 z_{IR}^2 ) \ (Q^2 + D^2) \rb ,
\eea
where the range of $D$ given by $0 \le | D | \le Q$.
If we set $D=0$, the above can reproduce the result in Ref. \cite{Lee:2009bya} with the redefinition 
of parameters. The Hawking-Page transition occurs at $\D S =0$. Suppose that at the critical point $z_h = z_c$, $\D S$ becomes zero. At this critical point describing the Hawking-Page transition, $Q^2 + D^2$ can be determined as a function of $z_c$ from \eq{eq:hp} 
\be		\la{rel:density}
Q^2 + D^2 = \frac{12 R^2 g^2}{ \k^2 z_{IR}^4 z_c^4} \ \frac{z_{IR}^4 - 2 z_c^4}{5 z_c^2 - 2 z_{IR}^2} . 
\ee
In terms of $z_c$, the temperature in \eq{def:temp} can be rewritten as
\be		\la{rel:temp}
T_{RN} \equiv \frac{1}{\b_{RN}} = \frac{1}{ \pi z_c} \ls 1 - \frac{z_c^2}{2 z_{IR}^4} \ 
\frac{z_{IR}^4 - 2 z_c^4}{5 z_c^2 - 2 z_{IR}^2} \rs .
\ee

\begin{figure}
\vspace{-2cm}
\centerline{\epsfig{file=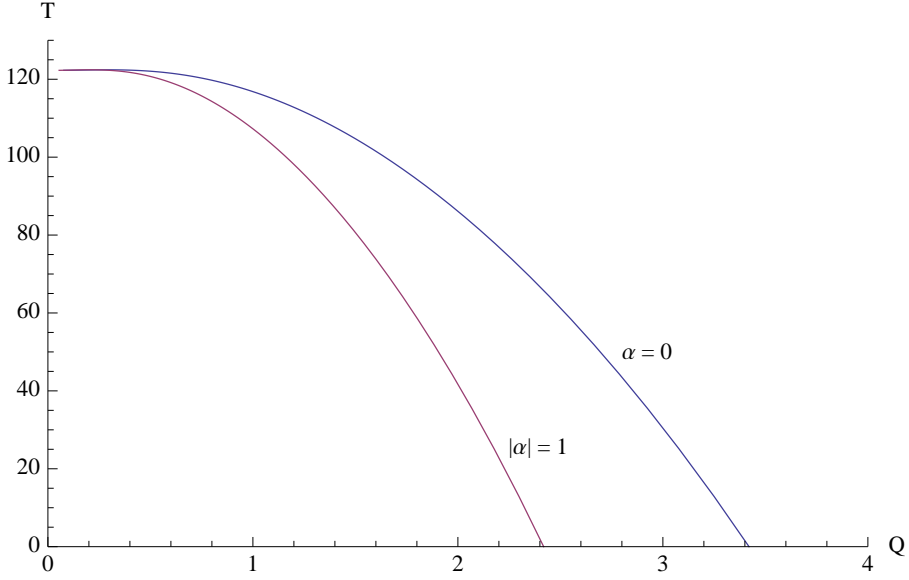,width=12cm}}
\vspace{0cm}
\caption{\small The Hawking-Page transition depending on the density difference between u- and d-quarks, where we set $R=1$ and $1/\k^2 = N_c^2/4 \pi^2 R^3$ with $N_c=3$ and $z_{IR} = 1/323 MeV$.}
\label{density}
\end{figure}

For a fixed $Q$, if we rewrite $D$ as $\a Q$ with $0 < |\a| < 1$, \eq{rel:density}
can be rewritten as
\be		\la{rel:den}
Q  = \sqrt{\frac{1}{1+ \a^2} \frac{12 R^2 g^2}{ \k^2 z_{IR}^4 z_c^4} \ \frac{z_{IR}^4 - 2 z_c^4}{5 z_c^2 - 2 z_{IR}^2}} . 
\ee
Using \eq{rel:temp} and \eq{rel:den}, we draw the Hawking-Page transition curve in Figure 1.
For $\a =0$, our system has equal number of u- and d-quarks and there is no symmetry energy.
In this case, the Hawking-Page
transition is the same as one obtained in Ref. \cite{Lee:2009bya} with different normalization.
For $\a =1$, the system is composed of only one species, u- or d-quark, and has the maximum
value of the symmetry energy. In Figure 1, the Hawking-Page transition temperature at a given $Q$ decreases due to the symmetry energy caused by the isospin.


\section{Conclusion}

We investigated the symmetry energy of the nuclear matter by using the hard wall model with 
$U(2)$ flavor group. This $U(2)$ flavor group contains two $U(1)$ diagonal subgroups, which are dual to quark density and isospin respectively. Using this fact, we considered the nuclear matter, proton and neutron. For the confining behavior, we introduced an IR cut off in the thermal charged AdS background. After calculating the free energy of this system, we found that the symmetry energy of the nuclear matter is proportional to the square of the nucleon density. 

We also calculated the free energy of quark-gluon plasma composed of u- and d- quarks. In this case, there exists a similar symmetry energy coming from the number difference of u- and d-quarks. In the high density and high temperature region, the symmetry energy of the quark matter is proportional to the square of the quark density and the inverse square of the temperature.

Finally, by comparing two free energies for the nuclear matters and the quark matter, we investigated the effect of the symmetry energy on the Hawking-Page transition corresponding to the deconfinement phase transition of the dual gauge theory. We found that at a given quark density, the symmetry energy decreases the Hawking-Page transition temperature.

\vspace{1cm}

{\bf Acknowledgement}

This work was supported by the National Research Foundation of Korea(NRF) grant funded by
the Korea government(MEST) through the Center for Quantum Spacetime(CQUeST) of Sogang
University with grant number 2005-0049409. C. Park was also
supported by Basic Science Research Program through the
National Research Foundation of Korea(NRF) funded by the Ministry of
Education, Science and Technology(2010-0022369).

\vspace{1cm}


\end{document}